\documentclass[preprint,showpacs,preprintnumbers,amsmath,amssymb]{revtex4}

\usepackage{graphicx}
\usepackage{dcolumn}
\usepackage{bm}

\begin{document}


\title{Some properties of the newly observed $X(1835)$ state at BES}

\author{Xiao-Gang He$^{1,2}$, Xue-Qian Li$^1$, Xiang Liu$^1$ and J.P. Ma$^3$}
\email{hexg@phys.ntu.edu.tw, lixq@nankai.edu.cn,
lx032278@phys.nankai.edu.cn, majp@itp.ac.cn}
 \affiliation{%
$^1$Department of Physics, Nankai University, Tianjin\\
$^2$NCTS/TPE, Department of Physics, National Taiwan University, Taipei\\
$^3$Institute of Theoretical Physics, Academia Sinica, Beijing
}%

\date{\today}

\begin{abstract}

{Recently the BES collaboration has announced observation of a
resonant state in the $\pi^+\pi^- \eta'$ spectrum in $J/\psi \to
\gamma \pi^+\pi^-\eta'$ decay. Fitting the data with a $0^{-+}$
state, the mass is determined to be 1833.7 MeV with $7.7\sigma$
statistic significance. This state is consistent with the one
extracted from previously reported $p \bar p$ threshold
enhancement data in $J/\psi \to \gamma p \bar p$. We study the
properties of this state using QCD anomaly and QCD sum rules
assuming $X(1835)$ to be a pseudoscalar and show that it is
consistent with data. We find that this state has a sizeable
matrix element $<0|G\tilde G|G_p>$ leading to branching ratios of
$(2.61\sim 7.37)\times 10^{-3}$ and $(2.21\sim 10.61)\times
10^{-2}$ for $J/\psi \to \gamma G_p$ and for $G_p \to \pi^+\pi^-
\eta'$, respectively. Combining the calculated branching ratio of
$J/\psi \to \gamma G_p$ and data on threshold enhancement in
$J/\psi \to \gamma p \bar p$, we determine the coupling for $G_p-
p-\bar p$ interaction. We finally study branching ratios of other
$J/\psi \to \gamma + \mbox{three mesons}$ decay modes. We find
that $J/\psi \to \gamma G_p \to \gamma (\pi^+\pi^- \eta, K K
\pi^0)$ can provide useful tests for the mechanism proposed. }

\end{abstract}

\maketitle

Recently the BES collaboration has announced observation of a
resonant state in the $\pi^+\pi^- \eta'$ spectrum\cite{etap} in
$J/\psi \to \gamma \pi^+\pi^- \eta'$.  A fit for a $0^{-+}$
resonant state with a Breit-Wigner function yields a mass $M =
1833.7\pm 6.5(stat)\pm 2.7(syst)$ MeV, a width $\Gamma = 67.7\pm
20.3(stat) \pm 7.7(syst)$ MeV and a production branching ratio
$B(J/\psi \to \gamma X) B(X\to \pi^+\pi^- \eta') = (2.2\pm
0.4(stat)\pm 0.4(syst))\times 10^{-4}$ with a $7.7\sigma$
statistic significance. The mass of this state is consistent with
that extracted from the enhanced threshold $p\bar p$ events
in\cite{pp} $J/\psi \to \gamma p \bar p$. The properties of this
state cannot be explained by known particles. Various models have
been proposed to explain the possible
resonance\cite{newwork0,newwork1,newwork2}. Further experimental
confirmation of this state is needed.

There are some hypothetic candidate particles which may fit in the
picture. Some of the possibilities include a $p\bar p$ bound
state\cite{newwork0,newwork1}, and a pseudoscalar glueball
state\cite{newwork0,newwork2}. The existence of glueballs is a
natural prediction of QCD. The prediction for the glueball masses
is, however, a non-trivial task. QCD sum rules\cite{gmass,hhh} and
lattice QCD\cite{lattice} calculations obtain the lowest
pseudoscalar glueball mass to be in the range of 1800 to 2600 MeV
with lattice calculations giving a mass in the upper range. One
cannot rule out the possibility that the resonant state $X(1835)$
is a glueball based on our present understanding of the glueball
masses alone. At this stage there is no compelling reasons to
believe that the resonance is a $p\bar p$ bound state
either\cite{newwork1}. These speculative particle states, although
attractive, their existences are far from being established. More
theoretical and experimental efforts are needed to go further. At
a more modest level, even to know whether the data from BES can be
consistently explained by a specific resonance and to further test
the mechanism, more information about properties of the resonance
is needed, such as how it is produced in radiative $J/\psi$ decays
and how it decays into other particles.

In this work we study the properties of the $X(1835)$ resonance
using QCD anomaly and QCD sum rules assuming that this state is a
pseudoscalar $G_p$ which couples strongly with two gluons which
may or may not be a glueball or a $p\bar p$ bound state depending
on whether this state has large mixing. We find that the matrix
element $<0|G\tilde G|G_p>$ is larger than $<0|G\tilde
G|\eta(\eta')>$ indicating a large glue content in $G_p$ which is
usually referred to as a glueball in the literature. This leads to
large branching ratios of $(2.61\sim 7.37)\times 10^{-3}$ for
$J/\psi \to \gamma G_p$ and $(2.21\sim 10.61)\times 10^{-2}$ for
$G_p \to \pi^+\pi^- \eta'$. The coupling for $G_p- p-\bar p$
interaction can also be determined. We finally discuss how other
$G_p$ decay modes can be used to test the mechanism.

There have been considerable amount of literatures on production
of a pseudoscalar in radiative $J/\psi$ decays, in terms of QCD
sum rules\cite{shifman} and perturbative QCD
calculations\cite{pqcd,close}. We follow the QCD sum rule approach
in Ref.\cite{shifman} such that we can treat radiative $J/\psi$
decays into $\eta$, $\eta'$ and $G_p$ in the same framework with
QCD anomaly. In this framework, the radiative $J/\psi$ decay
amplitudes are determined as follows: one first evaluates the
internal charm quark loop contribution to the interaction of
two-photon $\to$ two-gluon, and then saturates the $c\bar c$ pair
which couples to one of the external photons by $J/\psi$ and other
resonant states using the standard procedure of the QCD sum rules.
The two gluons are then converted into the related pseudoscalar
states. This approach works the best when the final pseudoscalar
has a mass-squared much smaller than $4m^2_c\sim m^2_{J/\psi}$.
For a pseudoscalar of mass 1833.7 MeV, there may be large
corrections from significant two-photon to multi-gluon couplings
since the factor $m^2_{G_p}/m^2_{J/\psi}$ may not be sufficient to
suppress higher order contributions. However, one expects that the
matrix elements of operators converting multi-gluon  to a
pseudoscalar $G_p$ must be smaller compared with that from the
leading two gluon operator. The situation may not be too severe to
damage the whole picture of the two gluon scenario. Also in our
later discussions we will only use the ratios of two different
$J/\psi \to \gamma X_i$ branching fractions, a large part of the
uncertainty is expected to be cancelled out. One expects that the
error range can be controlled to be within a factor of two.

In this calculation the two-gluon operator with appropriate
quantum numbers is, $G_{\mu\nu}\tilde G^{\mu\nu}$.  The matrix
elements converting the two gluons into a pseudoscalar $X_i$ is
usually parameterized as: $f_{i} m^2_{i} = \langle 0|(3\alpha_s/4
\pi)G_{\mu\nu} \tilde G^{\mu\nu}|X_i \rangle$. Since the rest of
the decay amplitude for $J/\psi \to \gamma X_i$ is independent of
the final pseudoscalar state, the ratio of radiative branching
fractions for $X_i$ and $X_j$ states is simply given
by\cite{shifman}
\begin{eqnarray}
R_{ij} = {B(J/\psi\to \gamma X_i)\over B(J/\psi \to \gamma X_j)} =
{|f_{i}m^2_{i}|^2\over |f_{j} m^2_{j}|^2}
{(1-m^2_{i}/m^2_{J/\psi})^3\over (1-m^2_{j}/m^2_{J/\psi})^3}.
\label{gamma}
\end{eqnarray}

The parameters $f_{\eta,\eta',G_p}$ play a crucial role in
determination of $J/\psi \to \gamma G_p$ in QCD sum rule approach.
The parameters $f_{\eta,\eta'}$ can be easily obtained from the
QCD anomaly relations in the limit that the strange quark mass is
much larger than the up and down quark masses. One
has\cite{anomaly}
\begin{eqnarray}
&&\langle 0|{3\alpha_s\over 4\pi}G_{\mu\nu} \tilde
G^{\mu\nu}|\eta\rangle = \sqrt{3\over 2} (\cos\theta f_8 -
\sqrt{2} \sin\theta f_0)
m^2_\eta,\nonumber\\
&&\langle 0|{3\alpha_s\over 4\pi}G_{\mu\nu} \tilde
G^{\mu\nu}|\eta'\rangle  = \sqrt{3\over 2} (\sin\theta f_8 +
\sqrt{2} \cos \theta f_0) m^2_{\eta'},
\end{eqnarray}
where $\theta$ is the $\eta$-$\eta'$ mixing angle with $\eta =
\eta_8 \cos\theta - \eta_0 \sin \theta$ and $\eta' = \eta_8 \sin
\theta + \eta_0 \cos\theta$. $f_{8,0}$ are decay constants of the
SU(3) octet $\eta_8$ and singlet $\eta_0$.

There are many theoretical studies on the values of $\theta$ and
$f_{8,0}$. In our study since only $\eta$, $\eta'$ and $J/\psi$
are involved, we will use related processes to determine $f_{8,0}$
and $\theta$. These processes include $\eta\to \gamma\gamma$,
$\eta'\to \gamma\gamma$ and $R_{\eta'\eta}= B(J/\psi\to \gamma
\eta')/B(J/\psi \to \gamma \eta)$. We have
\begin{eqnarray}
&&\Gamma(\eta\to \gamma\gamma) = {m^3_\eta\over 96
\pi^3}\alpha^3_{em} \left ( {\cos\theta\over f_8} -
2\sqrt{2}{\sin\theta\over f_0}\right )^2,\nonumber\\
&&\Gamma(\eta'\to \gamma\gamma) = {m^3_{\eta'}\over 96
\pi^3}\alpha^3_{em} \left ( {\sin\theta\over f_8} +
2\sqrt{2}{\cos\theta\over f_0}\right )^2,\\
&&R_{\eta'\eta} = \left \vert {(\sin\theta f_8 + \sqrt{2} \cos
\theta f_0) m^2_{\eta'}\over (\cos\theta f_8 - \sqrt{2} \sin\theta
f_0) m^2_\eta}\right \vert^2 {(1-m^2_{\eta'}/m^2_{J/\psi})^3\over
(1-m^2_{\eta}/m^2_{J/\psi})^3}.\nonumber
\end{eqnarray}

Using experimental values $B(\eta\to \gamma\gamma) =
(39.43\pm0.26)\%$, $B(\eta'\to \gamma\gamma)=(2.12\pm 0.14)\%$,
$B(J/\psi\to \gamma \eta)=(8.6\pm 0.8)\times 10^{-4}$ and
$B(J/\psi \to \gamma \eta')=(4.31\pm0.30)\times 10^{-3}$
\cite{PDG}, we obtain the ranges (central values) for the
parameters as: $\theta = -16.88^{\circ}\sim
-18.60^{\circ}$$(-17.72^{\circ})$, $f_8 = 0.98 f_\pi\sim
1.04f_\pi$ $(1.01f_\pi)$ and $f_0=1.06 f_\pi\sim 1.21f_\pi$
$(1.08f_\pi)$ with $f_\pi = 132$ MeV being the pion decay
constant. The correlations of these parameters are shown in Fig.
1. These values are consistent with the values determined from
other considerations\cite{other} for $\theta$ and $f_{8,0}$. This
gives us some confidence in using the QCD sum rule results for
$J/\psi \to \gamma \eta$, $J/\psi \to \gamma \eta'$, and as well
as for $J/\psi \to \gamma G_p$. We will use the above values for
$\theta$ and $f_{8,0}$ in our later discussions.

\begin{figure}[htb]
\begin{center}
\scalebox{0.6}{\includegraphics{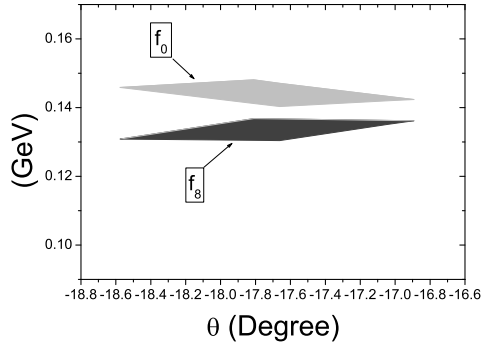}}
\end{center}
\caption{The dependence of $f_{0}$ and $f_{8}$ on $\theta$. The
ranges are due to one $\sigma$ errors of data points.}
\end{figure}

So far the parameter $f_{G_p}$ is not well understood. To obtain
some information, we use the QCD sum rules to calculate it. The
basic idea of QCD sum rule analysis in the present case is to
match the dispersion relation involving the hadronic spectral
density $\rho(s)$ to the vacuum topological susceptibility
$T(-q^2) = i \int d^4x e^{iq\cdot x} \langle 0 |T[j_{ps}(x)
j_{ps}(0)]|0\rangle$ with the result found by using the operator
product expansion. We follow Ref.\cite{hhh} by making a Borel
transformation on $T(s)$ with $\int^{\infty}_0 Im(T(s) e^{-s/M^2}
ds/s$ and $\int^{\infty}_0 Im(T(s)) e^{-s/M^2} ds$, and including
the two ground pseudoscalar states $\eta$ and $\eta'$, and $G_p$
in the resonant spectral density, to obtain the leading order
matching conditions\cite{hhh}
\begin{eqnarray}
&&f^2_{G_p}m^2_{G_p}e^{-m^2_{G_p}/M^2} + f^2_\eta
m^2_{\eta}e^{-m^2_\eta/M^2} + f^2_{\eta'}
m^2_{\eta'}e^{-m^2_{\eta'}/M^2}\nonumber\\
&&= \int^{s_1}_0 b s e^{-s/M^2}ds + \tilde D_4 + O({1\over M^2}) + inst. ,\nonumber\\
&&f^2_{G_p}m^4_{G_p}e^{-m^2_{G_p}/M^2} + f^2_\eta
m^4_{\eta}e^{-m^2_\eta/M^2} + f^4_{\eta'}
m^2_{\eta'}e^{-m^2_{\eta'}/M^2}\nonumber\\
&& = \int^{s_1}_0 b s^2 e^{-s/M^2} ds - \tilde D_6 + O({1\over
M^2}) + inst. , \label{sumrule}
\end{eqnarray}
where ``inst'' indicates direct instanton effects\cite{instanton}.
$\tilde D_{4,6}$ are related to the gluon condensations,
$D_4 = 4 \langle 0|G_{\mu\nu}G^{\mu\nu}|0\rangle$, and $D_6 = 8
g_sf_{abc} \langle 0|G^a_{\mu\alpha}G^{b,\alpha}_\nu
G^{c,\nu\mu}|0\rangle$, with $\tilde D_4 = \pi(3\alpha_s/4\pi)^2
D_4 $, $\tilde D_6 = \pi(3\alpha_s/4\pi)^2 D_6$. $b =
(3\alpha_s/4\pi)^2(2/\pi)(1+ 5 \alpha_s/\pi)$. In our numerical
evaluations, we will use $\langle 0|\alpha_{s}G^2|0\rangle=(7.1\pm
0.9)\times 10^{-2}$ GeV$^4$ and the relation $\langle 0|
g^{3}f_{abc}G^{a}G^{b}G^{c}|0\rangle=1.2\;\mathrm{GeV}^{2}\langle0|
\alpha_{s}G^{2}|0\rangle$~\cite{D46}. We comment that there are
other $0^{-+}$ states around 1400 MeV region which may contribute
to the spectrum density if these states contain large two gluon
contents. We will assume that the gluon contents are small in
these states and their contributions to the spectrum density can
be neglected.

To determine $f_{G_p}$, we take the usual practice to find the
parameters $f_{G_p}$ and $s_{1}$ for a given Borel parameter $M$
and look for a region where the dependence of $f_{G_p}$ and $s_1$
on $M$ is insensitive. We will negelct the direct instanton
effect in our calculation and will come back to comment on the effects
later.
Note that the analysis with a fixed $G_p$
mass here is different than previous ones\cite{gmass,hhh,D46}
where the mass of $G_p$ is taken as one of the parameters to be
determined and therefore the results is in general different. The
solutions for $f_{G_p}$ depend on the value $\alpha_s$ which we
take to be the value at the scale $\mu = m_{G_p}$ with $\alpha_s =
0.35\pm 0.05$. We find that solutions exist only for a restricted
parameter space for $\alpha_s$ and $\tilde D_{4,6}$. In certain
ranges, for a given set of input values of $\alpha_s$ and $\tilde
D_{4,6}$, there are two solutions. For example with $\alpha_s =
0.39$, $\tilde D_4=1.99\times 10^{-2}\;\mathrm{GeV}^{4}$ and
$\tilde D_6=3.79\times 10^{-3}\;\mathrm{GeV}^{6}$, we get the two
solutions to be:  a) $s_1=3.2$ GeV$^2$ and $f_{G_p}=0.081$ GeV,
and b) $s_1 = 3.5$ GeV$^2$ and $f_{G_p} = 0.091$ GeV. When $M$ is
larger than $7$ GeV, the solutions are fairly stable. As long as
we choose an $M$ far above $m_{G_p}$, the power corrections from
higher dimensional operators on the right hand sides of eq.
(\ref{sumrule}) can be suppressed.  We note that for the solution
with lower $s_1$, the value of $s_1$ is smaller than $m^2_{G_p}$
which cannot be considered to be a good solution since it implies
that the continuum already starts to contribute to the sum rules
in the resonant region in contradiction with the QCD sum rules
assumptions. We therefore should choose the solution with the
larger $s_1$. This solution allows a small gap between the
resonances and the continuum.  We show the results in Fig. 2
allowing $\alpha_s$ to vary from 0.3626  (where solution begins to
exist) to $0.4$ (the one $\sigma$ allowed upper bound), and all
other quantities, $\tilde D_{4,6}$, $f_{\eta,\eta'}$ and $\theta$
to vary within one $\sigma$ error ranges. We see that the
dependence on $M$ is very mild. We conclude that there are
consistent solutions from QCD sum rules for a pseudoscalar of mass
1833.7 MeV, and obtain a conservative range for $f_{G_p}$ with
\begin{eqnarray} f_{G_p} = 0.072\sim 0.100\;\mbox{GeV}.\label{fG}
\end{eqnarray}

\begin{figure}[htb]
\begin{center}
\scalebox{0.6}{\includegraphics{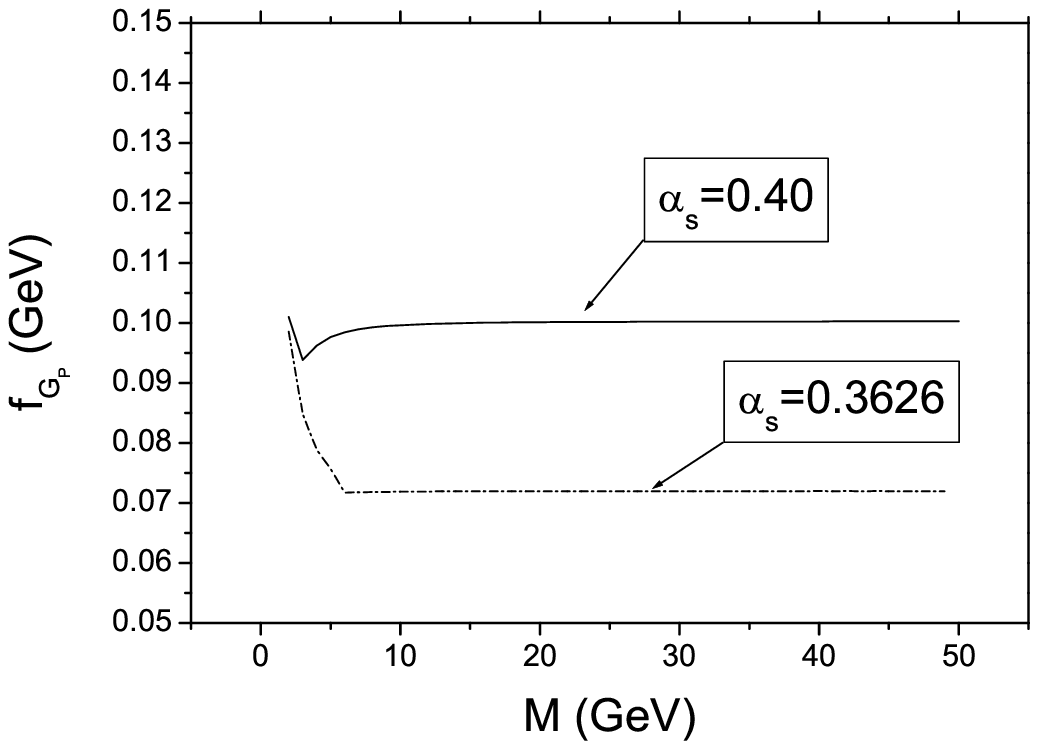}}\\\scalebox{0.6}{\includegraphics{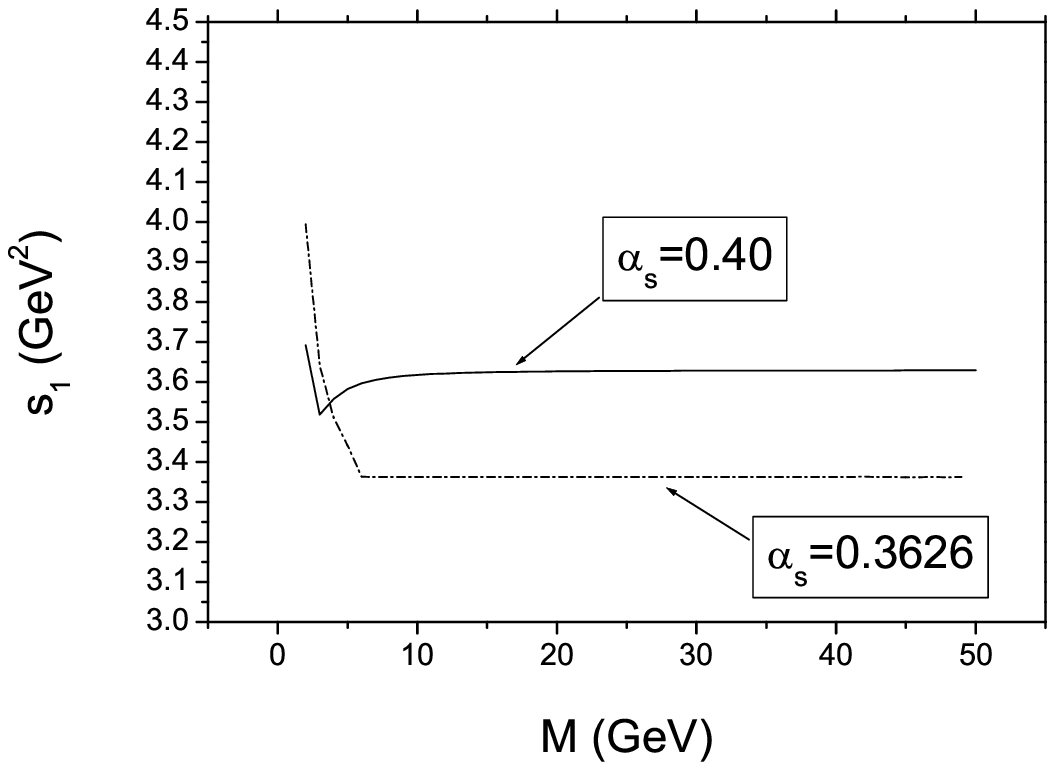}}
\end{center}
\caption{$f_{G_p}$ and $s_1$ as functions of $M$. The upper and
lower bounds are for one $\sigma$ ranges of $\tilde D_{4,6}$,
$f_{\eta,\eta'}$ and $\theta$ with $\alpha_s = 0.4$ and $\alpha_s
= 0.3626$, respectively.}
\end{figure}

When we obtained the range of $f_{G_p}$, the direct instanton
effects were neglected. At large $Q$, the instanton effects are
suppressed\cite{gmass,newwork2,suppression}, for example spike
distribution for instanton density results in an exponential
suppression when $Q^2$ becomes larger than a GeV$^2$ or
so\cite{gmass}. With the glueball mass fixed at 1835 MeV, the
suppression may not be sufficient to neglect the contributions
from the direct instanton. The effects of direct instanton may be
substantial. However the detailed calculations depend on the
instanton density and the density shape. A reliable evaluation of
the instanton effects is difficult. Nevertheless model
calculations show that instanton effects may be important due to
modification to the normalization of the Wilson coefficients for
the relevant operators. We will not go into the specific details
as it is too model dependent. Some detailed discussions for direct
instanton effects can be found in Ref.\cite{gmass}. We emphasize
that QCD sum rule results should be taken as an estimate within a
factor of two. In our later discussions we will use the range for
$f_{G_p}$ obtained the above as a reference. Should a more precise
value will be determined with some method, one can easily rescale
the values accordingly.

With the above range for $f_{G_p}$, we find that the matrix
element $<0|\alpha_s G\tilde G|G_p>$ is larger than $<0|\alpha_s
G\tilde G|\eta(\eta')>$ indicating that $G_p$ contains a large
gluon content. With $f_{G_p}$ determined, we are now able to
obtain information on the range for $B(J/\psi \to \gamma G_p)$
combining eq. (\ref{gamma}) and experimental data on $J/\psi \to
\gamma \eta(\eta')$. We have
\begin{eqnarray}
B(J/\psi \to \gamma G_p) =(2.61\sim 7.37)\times 10^{-3}.
\end{eqnarray}

Using the BES data $B(J/\psi \to \gamma G_p) B(G_p\to \pi^+\pi^-
\eta') = (2.2\pm 0.4(stat)\pm 0.4(syst))\times 10^{-4}$, we can
therefore also obtain the branching ratio of $G_p$ to $\pi^+\pi^-
\eta'$ using our estimate of $J/\psi \to \gamma G_p$. We have
\begin{eqnarray}
B(G_p\to \pi^+\pi^- \eta') = (2.21\sim 10.61)\times
10^{-2}.\label{mesonb}
\end{eqnarray}
The branching ratio for this decay is large, but it does not
contradict with known data.

If the enhanced threshold $p\bar p$ events in $J/\psi \to \gamma p
\bar p$ is also due to the same $G_p$ state, we can obtain
information about the interaction of $G_p$ with a proton and
anti-proton pair, $L = C_{GB} \bar p \gamma_5 p G_p$. Since the
mass of $G_p$ is slightly below the threshold of two proton mass
$2m_p$, one cannot simply take $B(J/\psi \to \gamma G_p \to \gamma
p \bar p)$ to be equal to $B(J/\psi \to \gamma G_p)B(G_p\to p \bar
p)$. One must consider the off-shell effects of $G_p$ in terms of
the Breit-Wigner approach. We have
\begin{eqnarray}
&&B(J/\psi\to \gamma G_p \to \gamma p \bar p)  = C^2_{GB}
{B(J/\psi \to \gamma G_p)\over 8 \pi^2}\\
&&\times \int^{m^2_{J/\psi}}_{4m^2_p}dq^2 {(1-q^2/m^2_{J/\psi})^3
q^2\over (1-m^2_{G_p}/m^2_{J/\psi})^3}{\sqrt{1-4m^2_p/q^2}\over
(q^2-m^2_{G_p})^2 + \Gamma_{G_p}^2 m^2_{G_p}}.\nonumber
\end{eqnarray}
In the above, we have assumed that $C_{GB}$ does not depend on
$q^2$ sensitively and has been moved out from the integration
sign.

Experimental data on $B(J/\psi\to \gamma G_p\to \gamma p\bar p) =
(7.0 \pm 0.4^{+1.8}_{-0.8})\times 10^{-5}$ then implies
\begin{eqnarray}
C_{GB} = 0.42\sim 0.85.
\end{eqnarray}
One sees that threshold enhancement data in $J/\psi \to \gamma p
\bar p$ can also be consistently explained.

We now discuss how the resonance $G_p$ may decay into other
particles. As have been pointed out earlier this state has a large
gluon contents, it may be a glueball which is a SU(3) singlet. Of
course one needs to be more open minded that it may has sizeable
mixing with other state. We will take the state $G_p$ to be a
glueball state and study the consequences from flavor symmetry
point of view. One can easily study the branching ratios for
$J/\psi \to \gamma G_p\to \gamma B \bar B$ with $B \bar B$ a pair
of octet baryons. The mixing effect can be easily implemented by
introducing some mixing parameters.

The coupling of a SU(3) singlet $G_p$ and octet baryon can be
written as $L= C_{GB} G_p Tr \bar B \gamma_5 B$ with SU(3) flavor
symmetry. In Table 1 we list the ratios of $r(B \bar B) = B(J/\psi
\to \gamma G_p \to \gamma B \bar B)/B(J/\psi \to \gamma G_p \to
\gamma p \bar p)$ for possible decay modes. We comment that the
$X(1835)$ contributions listed in Table 1 hold as long as the
resonance is an SU(3) singlet and does not depend on the size of
$C_{GB}$. If the resonance transforms non-trivially under the
flavor SU(3) symmetry, the predictions would be
different\cite{helima}. In principle experimental measurements of
these branching ratios can provide important information about the
nature of the resonance. However, the branching ratios for other
baryon pair decay modes are much smaller than the branching ratio
with a proton and anti-proton pair except the neutron and
anti-neutron pair decay mode which is then experimentally
difficult to carry out. The usefulness of these decay modes
depends on whether, near the resonance region, the resonance
contributions dominate over the non-resonance continuum parts
which we will comment on later.

\begin{table}[htb]
\begin{center}
\begin{tabular}{|c|c|} \hline
$r(\Xi^0\bar \Xi^0)$&$1.58\times 10^{-3}$\\\hline $r(\Xi^-\bar
\Xi^-)$&$1.38\times 10^{-3}$\\\hline $r(\Sigma^+\bar
\Sigma^+)$&$1.45\times 10^{-2}$\\\hline $r(\Sigma^0\bar
\Sigma^0)$&$1.38\times 10^{-2}$\\\hline $r(\Sigma^-\bar
\Sigma^-)$&$1.28\times 10^{-2}$\\\hline $r(\Lambda\bar
\Lambda)$&$4.23\times 10^{-2}$\\\hline $r(n\bar n)$&$0.96$\\\hline
\end{tabular}
\caption{$r(B\bar B)$ for different $J/\psi \to \gamma G_P \to
\gamma B \bar B$ decays.}
\end{center}
\end{table}

A better test of the mechanism may come from other three meson
decay modes of $G_p$. To have further information on the branching
ratios for $G_p$ decay into three meson modes, we here follow Ref.
\cite{neufeld} to use U(3) chiral theory to describe how it
couples to known meson particles. Notice that the use of U(3)
symmetry will not change our previous discussions on $G_p \to B
\bar B$ results. The reason we use U(3) chiral perturbation theory
 for the interaction is that it can naturally include many properties of
 chiral anomaly which our calculations for $f_{i}$ depend on.
 To the leading order there are four terms which
may cause $G_p$ to decay\cite{neufeld}
\begin{eqnarray}
L &=& i a_1 \partial_\mu G_p  Tr\left (\Sigma^\dagger \partial^\mu
\Sigma
\partial_\nu \Sigma^\dagger \partial^\nu \Sigma \right )\nonumber\\
 &+& ia_2 G_p Tr\left (
(\Sigma^\dagger \partial^2 \Sigma - \partial^2 \Sigma^\dagger
\Sigma) \partial_\nu \Sigma^\dagger \partial^\nu
\Sigma \right )\nonumber\\
&+& ia_3 G_p Tr\left (\chi \Sigma - \Sigma^\dagger \chi\right
)\nonumber\\
&+& i a_4 G_p Tr\left (\chi \Sigma^\dagger \partial_\nu \Sigma
\partial^\nu \Sigma^\dagger - \chi \Sigma \partial_\nu
\Sigma^\dagger \partial^\nu \Sigma\right ),\label{ge}
\end{eqnarray}
where $\Sigma = exp[-i \sqrt{2} M/f]$ with $f = f_\pi/\sqrt{2}$
and $M$ is the U(3) meson nonet. $\chi$ is proportional to the
light quark masses and is given by $\chi = diag(m^2_\pi, m^2_\pi,
2 m^2_K - m^2_\pi)$. The last two terms in eq. (\ref{ge}) comes
from explicit U(3) (and SU(3)) breaking due to quark masses.

The two SU(3) breaking terms in eq. (\ref{ge}), if dominant, will
lead to the main decay mode to be $G_p\to \pi K K$\cite{neufeld}
with a very suppressed rate for $G_p \to \pi^+\pi^-\eta'$. The BES
data indicates that the decay mode $G_p \to \pi^+\pi^- \eta'$ has
a large branching ratio compared with other three particle decays
(yet to be discovered), therefore, these two terms may be
suppressed. If the first two terms dominate, we obtain the
effective Lagrangian for the decay amplitude $G_p \to \pi^+\pi^-
\eta'$ to be,
\begin{eqnarray}
L &=& {4\over \sqrt{3} f^3} (\sqrt{2} \cos\theta +\sin
\theta)\left ( a_1 \partial_\mu G_p (\partial^\mu \eta'
\partial_\nu \pi^+\partial^\nu \pi^-\right .\nonumber\\
& +& \left .\partial^\mu \pi^-
\partial_\nu \pi^+\partial^\nu \eta' + \partial^\mu \pi^+
\partial_\nu \eta'
\partial^\nu \pi^-)\right .\nonumber\\
&+& \left . a_2 G_p (\partial^2 \eta' \partial_\nu \pi^+
\partial^\nu \pi^- + \partial^2 \pi^- \partial_\nu \pi^+
\partial^\nu \eta'\right .\nonumber\\
& +&\left . \partial^2 \pi^+ \partial_\nu \eta'
\partial^\nu \pi^-)\right ).
\end{eqnarray}
which leads to a decay amplitude,
\begin{eqnarray}
&&M(G_p \to \pi^+\pi^- \eta')= {\sqrt{2} \cos\theta +
\sin\theta\over \sqrt{3} f^3}\nonumber\\
&&\times \left (  a_1 [(s-m^2_{G_p} -
m^2_{\eta'})(s-2m^2_\pi)\right .\nonumber\\
&&\left . + (t-m^2_{G_p} - m^2_\pi)(t-m^2_{\eta'}-m^2_\pi)\right
.\nonumber\\&& \left . +
(u-m^2_{G_p}-m^2_\pi)(u-m^2_{\eta'}-m^2_\pi)]\right .\nonumber\\
&&\left . + 2a_2[m^2_{\eta'}(s-2m^2_\pi)+ m^2_\pi
(t-m^2_{\eta'} - m^2_\pi)\right .\nonumber\\
&&\left . + m^2_\pi (u - m^2_{\eta'} - m^2_\pi)]\right ),
\end{eqnarray}
where $s= (p_{\pi^+}+p_{\pi^-})^2$, $t= (p_{\pi^-} + p_{\eta'})^2$
and $u = (p_{\pi^+}+p_{\eta'})^2$.

To obtain the BES result for $J/\psi \to \gamma \pi^+\pi^-\eta'$
with the  $B(J/\psi \to \gamma G_p)$ obtained earlier, we have
\begin{eqnarray}
|a_1|^2 - 0.376 Re(a_1 a_2^*) + 0.038|a_2|^2 = (0.94\sim 4.52)
\times 10^{-5} \mbox{GeV}^{-2}.
\end{eqnarray}

Since there are two parameters, $a_1$ and $a_2$, involved, with
data from $J/\psi \to \gamma G_p \to \gamma \pi^+\pi^- \eta'$
alone it is not possible to predict branching ratios for other
three meson decay modes. There are several kinematically allowed
decay modes which can provide more information. These additional
decay modes include: $\eta'\pi^{0}\pi^{0}$, $\eta\pi^{+}\pi^{-}$,
$\eta\pi^{0}\pi^{0}$, $\eta K^{+}K^{-}$, $\eta K^{0}\bar{K}^{0}$,
$\pi^{0}K^{0}\bar{K}^{0}$, $\pi^{0}K^{+}K^{-}$,
 $\pi^{+}K^{0}K^{-}$, $\pi^{-}\bar{K}^{0}K^{+}$, and $\eta\eta\eta$.
With the assumption that the dominant contributions come from the
$a_{1,2}$ terms, even with SU(3) breaking effects from the final
meson mass differences, one would have the following relations
between branching ratios: $B(\pi^0\pi^0 \eta'(\eta)) =
B(\pi^+\pi^- \eta' (\eta))/2$, $B(K^+K^- \eta) = B(K^0 \bar K^0
\eta)$, and $B(\pi^0 K^+K^-) = B(\pi^0 K^0 \bar K^0) = B(\pi^- K^+
\bar K^0)/2 = B(\pi^+ K^- K^0)/2$. We list the ratios
$r(p_1p_2p_3) = B(G_p\to p_1 p_2p_3)/B(G_p\to \pi^+\pi^- \eta')$
in Table 2. The branching ratio for $G_p \to \pi^+\pi^- \eta$ can
be larger than that for $G_p \to \pi^+\pi^- \eta'$. No observation
of $J/\psi \to \gamma \pi^+\pi^- \eta$ near the resonance would
imply that there may be large cancellations between terms
proportional to $a_1$ and $a_2$ for this decay mode. $G_p \to \pi
K K$ are some another modes which may have large branching ratios
near the resonance. The branching ratios here are different from
those predicted if the resonance transforms non-trivially under
the SU(3) symmetry\cite{newwork1}. These new decay modes can serve
to test the mechanism proposed in this paper. We urge the BES
collaboration to carry out an systematic analysis to obtain more
information about the properties of the resonant state $X(1835)$.

\begin{table}[htb]
\begin{center}
\begin{tabular}{|c|c|} \hline
$r(\pi^+\pi^-\eta)$&$11.26(|a_1|^2 -0.186Re(a_1a_2^*) + 0.009
|a_2|^2)/b(\pi^+\pi^-\eta')$\\\hline $r(K^+ K^-\eta)$&$1.57\times
10^{-3}(|a_1|^2 - 0.474Re(a_1a_2^*) + 0.056
|a_2|^2)/b(\pi^+\pi^-\eta')$\\\hline
$r(K^+K^-\pi^0)$&$2.47(|a_1|^2 - 0.291Re(a_1a_2^*) + 0.021
|a_2|^2)/b(\pi^+\pi^-\eta')$\\\hline
$r(\eta\eta\eta)$&$0.006(|a_1|^2 - 0.538Re(a_1a_2^*) + 0.073
|a_2|^2)/b(\pi^+\pi^-\eta')$\\\hline
\end{tabular}
\caption{$r(p_1p_2p_3)$ for various $G_p \to p_1p_2p_3$ decay
modes. Here $b(\pi^+\pi^- \eta') = |a_1|^2- 0.376 Re(a_1a_2^*) +
0.038 |a_2|^2$.}
\end{center}
\end{table}

In the above discussions about predictions for other decay modes,
only contributions from the resonance are included. There are also
non-resonance effects. To isolate the resonance contributions near
the resonance region, one should remove the off-resonance
contributions by extrapolating the data away the resonance to the
resonance region. Theoretical study of the non-resonance
contributions is more complicated. We briefly discuss how the
non-resonance contributions for $J/\psi \to \gamma B \bar B$ and
$J/\psi \to \gamma \pi^+\pi^- \eta(\eta')$ can be parameterized
using SU(3) flavor symmetry. The general form for non-resonance
$J/\psi \to \gamma B \bar B$ has been discussed in
Ref.\cite{helima}. As far as the SU(3) structure is concerned,
that is, neglecting Lorentz structure, we have the amplitudes for
non-resonance contributions to  $J/\psi \to \gamma B\bar B$ and
$J/\psi \to p_1 p_2p_3$ to be given by
\begin{eqnarray}
&&M(J/\psi \to \gamma B \bar B) \sim Tr[ \bar B ( D_\gamma \{Q,
B\} +
F_\gamma [Q, B])],\nonumber\\
&& M(J/\psi \to \gamma p_1p_2p_3) \sim \tilde a \eta_1 Tr(QM^2) +
+ \tilde b \eta_1 \eta_1 Tr(QM)  + \tilde c Tr(QM^3).
\end{eqnarray}
where $Q = diag(2/3,-1/3,-1/3)$ is the electric charge matrix.
$D_\gamma$, $F_\gamma$, $\tilde a$, $\tilde b$ and $\tilde c$ are
form factors.

The complications in extracting information on the form factors
$D_\gamma$ and $F_{\gamma}$, and, $\tilde a$, $\tilde b$ and
$\tilde c$ are two folds. One of them is that the Lorentz
structure is much more complicated, that is for each of the form
factors $D_{\gamma}$, $F_\gamma$, $\tilde a$ and $\tilde b$, there
are actually several of them depending on how the Dirac matrices
are inserted in the $B \bar B$ bi-baryon product, and how
derivatives are taken on the meson and baryon fields. Another is
that the form factors in general depend on combinations of the
invariant masses of pairs of particles in the final state.
Nevertheless, fitting data, one can obtain information about the
parameters. Measurements of branching ratios of $J/\psi \to \gamma
B\bar B, \gamma p_1p_2p_3$ via non-resonance can also provide
important information for understanding the whole physics picture.

In summary, we have studied some implications of a possible
$0^{-+}$ resonance in the $\pi^+\pi^-\eta'$ spectrum in $J/\psi
\to \gamma \pi^+\pi^- \eta'$ observed by BES. We have shown that
it has a sizeable matrix element for $<0|G\tilde G|G_p>$. The
branching ratios for $J/\psi \to \gamma G_p$ and $G_p \to
\pi^+\pi^-\eta'$, using QCD anomaly and QCD sum rules, are
determined to be
 $(2.61\sim 7.37)\times 10^{-3}$ and
$(2.21\sim 10.61)\times 10^{-2}$, respectively. The coupling for
$G_p- p-\bar p$ interaction is also determined. We conclude that a
pseudoscalar $0^{-+}$ with large gluon content can consistently
explain the data. We have also studied branching ratios of other
decay modes using SU(3) flavor symmetry. We find that $J/\psi \to
\gamma G_p \to \gamma (\pi^+\pi^- \eta, K K \pi^0)$ can provide
useful tests for the mechanism proposed here.

\noindent {\bf Acknowledgments} We thank N. Kochelev and S.
Narison for useful discussions. X.G.H thanks B. McKellar for
hospitality at the University of Melbourne while part of this work
was carried out. This work was supported in part by grants from
NSC and NNSFC (10421003).


\begin{thebibliography}{99}

\bibitem{etap} M. Ablikim et al., BES Collaboration,
Phys. Rev. Lett. {\bf 95}, 262001 (2005).

\bibitem{pp} J. Z. Bai et al., BES Collaboration, Phys. Rev. Lett.
{\bf 91}, 022001 (2003).

\bibitem{newwork0} For a recent reveiw see J. Rosner,
hep-ph/0508155.

\bibitem{newwork1} J. Rosner, Phys. Rev. {\bf D 68}, 014004 (2003);
A. Datta and P. O'connell, Phys. Lett. {\bf B 567}, 273 (2003); B.S.
Zou and H.C. Chiang, Phys. Rev. {\bf D 69}, 034004 (2004); X. Liu et
al., High Energy Phys. Nucl. Phys. {\bf 30}, 1 (2006); C.H. Chang
and H.R. Pang, Commun. Theor. Phys. {\bf 43}, 275 (2005); A.
Sibirtsev et al., Phys. Rev. {\bf D 71}, 054010 (2005); H. Loiseau
and S. Wyceck, Phys. Rev. {\bf C 72}, 011001 (2005); M.L. Yan et
al., Phys. Rev. {\bf D 72}, 034027 (2005); G.J. Ding and M.L. Yan,
Phys. Rev. {\bf C 72}, 015208 (2005); S.L. Zhu and C.S. Gao, Commun.
Theor. Phys. {\bf 46}, 291 (2006).

\bibitem{newwork2} N. Kochelev and D.P. Min, Phys. Lett. {\bf B 633}, 283 (2006).

\bibitem{gmass} S. Narison,
Nucl. Phys. {\bf B 509}, 312 (1998); H. Forkel, Phys. Rev. {\bf
D71}, 054008 (2005).

\bibitem{hhh} G. Gabadadze, Phys. Rev. {\bf D 58}, 055003 (1998);
X.G. He, W.S. Hou and C.S. Huang, Phys. Lett. {\bf B 429}, 99
(1998).

\bibitem{lattice} C.J. Morningstar and M. Peardon,
Phys. Rev. {\bf D 60}, 034509 (1999);
  A. Hart and M. Teper, Phys. Rev. {\bf D 65},
34502 (2002).

\bibitem{shifman}
V. A. Novikov, M. A. Shifman, A. I. Vainshtein and V. I. Zakharov,
Nucl. Phys. {\bf B 165}, 55 (1979).

\bibitem{pqcd} J.P. Ma, Nucl. Phys. {\bf B 605}, 625 (2001), Erratum-ibid, {\bf B611}, 523 (2001);
Phys. Rev. {\bf D 65}, 097506 (2002); X.G. He, H.Y. Jin and J.P. Ma,
Phys. Rev. {\bf D 66}, 074015 (2002).

\bibitem{close} F. Close, G. Farrar and Z.P. Li, Phys. Rev. {\bf
D 55}, 5749 (1997).

\bibitem{anomaly} P. Ball, J.M. Frere and M. Tytgat, Phys. Lett. {\bf B
365}, 367 (1996); R. Akhoury, J.M. Frere, Phys. Lett. {\bf B 220},
258 (1989); K.T. Chao, Nucl. Phys. {\bf B 317}, 597 (1989).



\bibitem{PDG} S. Eidelman et al., Particle Data Group, Phys. Lett.
{\bf B 592}, 1 (2004).

\bibitem{other} P. Ball, J.M. Frere and M. Tytgat in
\cite{anomaly}; A. Bramon, R. Ecribano and M.D. Scadron, Eur. Phys.
J. {\bf C 7}, 271 (1999).

\bibitem{instanton} V.A. Novikov et al., Nucl. Phys. {\bf B 191}, 301 (1981).

\bibitem{D46}
T. Huang, H.Y. Jin and A.L. Zhang, Phys. Rev. {\bf D 59}, 4026
(1999) ; S. Narison, Nucl. Phys. {\bf B 502}, 312 (1998).

\bibitem{suppression} V. A. Novikov et al., Phys. Lett. {\bf B 86}, 347 (1979).

\bibitem{helima} X.G. He, X.Q. Li and J.P. Ma, Phys. Rev. {\bf
D 71}, 014031 (2005).

\bibitem{neufeld} G. Gounaris and H. Neufeld, Phys. Lett. {\bf
B213}, 541 (1988), Erratum-ibid. {\bf B 218}, 508 (1989).



\end{thebibliography}
\end{document}